\documentclass[sigconf]{acmart}
\acmConference[MSR 2022]{MSR '22: Proceedings of the 19th International Conference on Mining Software Repositories}{May 23–24, 2022}{Pittsburgh, PA, USA}

\usepackage{color}
\usepackage{cite}

\usepackage{paralist} 
\usepackage{caption}
\usepackage{multirow}
\usepackage{booktabs,threeparttable}
\usepackage{color}
\usepackage{threeparttable}
\usepackage{colortbl}
\usepackage[autostyle=true,english=british]{csquotes}
\usepackage{graphicx}
\usepackage{subcaption}
\usepackage{url}
\usepackage{fancyhdr}
\usepackage{multirow}
\usepackage{hyperref}
\usepackage{tabularx} 
\usepackage{makecell}  
\usepackage{array}
\newcolumntype{C}{>{\centering\let\newline\\\arraybackslash\hspace{0pt}}X}
\newcolumntype{L}{>{\raggedleft\let\newline\\\arraybackslash\hspace{0pt}}X}
\newcolumntype{R}{>{\raggedright\let\newline\\\arraybackslash\hspace{0pt}}X}

\definecolor{hellgrau}{rgb}{0.95,0.95,0.95}
\definecolor{white}{rgb}{1,1,1}

\newcommand\rel{\textit{Relate}}
\newcommand\sub{\textit{Subtask}}
\newcommand\dup{\textit{Duplicate}}
\newcommand\clo{\textit{Clone}}
\newcommand\dep{\textit{Depend}}
\newcommand\epi{\textit{Epic}}
\newcommand\spl{\textit{Split}}
\newcommand\blo{\textit{Block}}
\newcommand\inc{\textit{Incorporate}}
\newcommand\cau{\textit{Cause}}

\newcommand\rep{\textit{Replace}}

\newcommand\Rel{\textit{General Relation}}
\newcommand\Dup{\textit{Duplication}}
\newcommand\Wor{\textit{Workflow}}
\newcommand\Tem{\textit{Temporal / Causal}}
\newcommand\Com{\textit{Composition}}

\newcommand\NL{\textit{Non-Links}}
\newcommand\OL{\textit{Other-Links}}

\copyrightyear{2022} 
\acmYear{2022} 
\setcopyright{acmlicensed}\acmConference[MSR '22]{19th International Conference on Mining Software Repositories}{May 23--24, 2022}{Pittsburgh, PA, USA}
\acmBooktitle{19th International Conference on Mining Software Repositories (MSR '22), May 23--24, 2022, Pittsburgh, PA, USA}
\acmPrice{15.00}
\acmDOI{10.1145/3524842.3528457}
\acmISBN{978-1-4503-9303-4/22/05}

\begin{document}
\title{Beyond Duplicates: Towards Understanding and Predicting Link Types in Issue Tracking Systems}

\author{Clara Marie L{\"u}ders}
\email{clara.marie.lueders@uni-hamburg.de}
\affiliation{%
  \institution{University of Hamburg}
  \city{Hamburg}
  \country{Germany}
}

\author{Abir Bouraffa}
\email{abir.bouraffa@uni-hamburg.de}
\affiliation{%
  \institution{University of Hamburg}
  \city{Hamburg}
  \country{Germany}
}

\author{Walid Maalej}
\email{walid.maalej@uni-hamburg.de}
\affiliation{%
  \institution{University of Hamburg}
  \city{Hamburg}
  \country{Germany}
}

\begin{abstract}
Software projects use Issue Tracking Systems (ITS) like JIRA to track issues and organize the workflows around them.
Issues are often inter-connected via different links such as the default JIRA link types \dup{},  \rel{}, \blo{}, or \sub{}.  
While previous research has mostly focused on analyzing and predicting duplication links, this work aims at understanding the various other link types, their prevalence, and characteristics towards a more reliable link type prediction.
For this, we studied 607,208 links connecting 698,790 issues in 15 public JIRA repositories.
Besides the default types, the custom types \dep, \inc, \spl, and \cau{} were also common. 
We manually grouped all 75 link types used in the repositories into five general categories: \Rel, \Dup, \Com, \Tem, and \Wor. Comparing the structures of the corresponding graphs, we observed several trends. 
For instance, \Dup{} links tend to represent simpler issue graphs often with two components and \Com{} links present the highest amount of hierarchical tree structures (97.7\%). 
Surprisingly, \Rel{} links have a significantly higher transitivity score than \Dup{} and \Tem{} links. 

Motivated by the differences between the link types and by their popularity, we evaluated the robustness of two state-of-the-art duplicate detection approaches from the literature on the JIRA dataset.  
We found that current deep-learning approaches confuse between \Dup{} and other links in almost all repositories.
On average, the classification accuracy dropped by 6\% for one approach and 12\% for the other. 
Extending the training sets with other link types seems to partly solve this issue. We discuss our findings and their implications for research and practice. 

\end{abstract}

\maketitle

\section{Introduction}

Development teams use Issue Tracking Systems (ITS) such as Bugzilla, Github Issues, or JIRA to track issues, including bugs to be fixed or features to be implemented. 
Over the years ITS have emerged as a central tool for planning and organizing development work~\citep{Zimmermann:TSE:2010}, and for communicating with users and other stakeholders~\citep{Bertram:CSCW:2010}.
However, a considerable pain point for stakeholders using ITS, as highlighted in the work of Fucci et al.~\citep{Fucci:ESEM:2018}, is the identification of dependencies between issues. 
Research found that linking issues can help reduce issue resolution time~\citep{Li:APSEC:2018} and prevent software defects~\citep{Rempel:IEEE:2017}.
Missing or incorrect links are problematic for requirements analysis and release planning~\citep{Thompson:MSR:2016}.
For instance, missing \dep{} or \blo{} links of an issue might be crucial for the assigned release. Missing \dup{} links might lead to missing information relevant for testing.  
A project might easily get thousands of issues. Each new issue might thus have hundreds of thousands of potentially relevant links. Correctly identifying and connecting issues quickly becomes difficult, time-consuming, and error-prone~\citep{Fucci:ESEM:2018, Lucassen:REW:2017}.

Most ITS allow the creation of different links between the issues to indicate technical or workflow dependencies.
For instance, Bugzilla allows to set properties such as ``depends on'', ``blocks'', and  ``See also'' for bug reports as well as to set the resolution status as ``duplicate'' with link to a duplicate report~\citep{Amoui:MSR:2013}. 
Similarly, JIRA users can choose between four possible default types in the \textit{Issue Links} section: \rel, \dup, \blo, and \clo\footnote{JIRA users can clone an issue. A \clo{} link is then automatically created.}. 
Additionally, \textit{Subtasks} and \textit{Epics} can be linked in separate sections. 
Organizations can create and use additional link types to meet their specific needs.
For instance, Qt\footnote{\url{https://bugreports.qt.io/secure/Dashboard.jspa}} uses~6 link types including \spl{} and \rep{}. Apache\footnote{\url{https://issues.apache.org/jira/secure/Dashboard.jspa}} uses as many as~21 including the custom link types \textit{Container} or \textit{Breaks}.
Each link type usually has an explicit definition in the ITS. 
Over time, stakeholders might also develop an ``implicit'' understanding of the connection represented by the link type. 
This might be either indeed unique to the community or simply a different name denoting the same connection labeled differently in another community. 
For instance, Apache uses both the link type \dep{} and \blo{}, while all other repositories only use one of these types predominantly. 
In Bugzilla these two link types are equivalent\footnote{\url{https://bugzilla.readthedocs.io/en/latest/using/understanding.html}}.

Studying the various link types and their usage patterns across the communities is essential for supporting issue linking and tightening automated tool support particularly predicting missing links \citep{Robillard:2014, Zhang:CJ:2016} to alleviate the burden of dependency identification for stakeholders~\citep{Fucci:ESEM:2018}.
In recent years, research has intensively studied the specific type \dup{}. 
Detecting those links would reduce the resolution time of duplicated issues and might reveal additional information included in one but not the other issue~\citep{Cavalcanti:CSMR:2010, Bettenburg:ICSM:2008}. 
Based on a Bugzilla dataset by Lazar et al.~\citep{Lazar:MSR:2014}, researchers recently presented duplicate prediction approaches using state-of-the-art machine learning models with top performances of up to~97\%~\citep{Deshmukh:ICSME:2017, He:ICPC:2020}. 

This work takes a holistic view on issue link types. 
We report on a study comparing the \textit{various types} and their usage in~15 well-known public JIRA repositories~\citep{Montgomery:2022}. 
By studying link types in JIRA, a widely used ITS in practice, we hope to create awareness about how the other types beyond duplicates are used and to inform a more generalizable and reliable link type predictions.
Our work has three specific contributions.
First, we manually reviewed and analyzed all
 types found in the ITS and categorized them into five general link categories to ease  comparison. 
We report on the types, categories, and usage frequencies across the studied repositories. 
Second, we apply techniques from the field of graph theory to compare the complexity, shape,  transitivity~\citep{Schank:2005}, and assortativity~\citep{Newman:APS:2002} of the different issue graphs corresponding to each link category. 
Our comparison reveal structural similarities across the repositories and several expected and unexpected trends. 
Third, we show that current link prediction models seem to rather learn the existence of the links instead of the specific link types. 
We discuss how future research should deal with this to improve prediction reliability.
We also share our analysis code and the labeled data to ease replication.\footnote{\url{https://github.com/RegenKordel/LYNX-BeyondDuplicates}}

The remainder of the paper is structured as follows. 
Section~\ref{sec:research-method} outlines our research questions, method, and data. 
Section~\ref{sec:link-types} reports on the link types and how they are used across the repositories . 
Section~\ref{sec:link-prediction} presents the results of our link prediction experiments based on state-of-the-art duplicate detection approaches.
In Section~\ref{sec:discussion}, we discuss how our results can be used by practitioners to monitor and tune their link usage and by researchers to develop and evaluate more precise link prediction approaches.
Finally, we discuss related work in Section~\ref{sec:rel-work} and conclude the paper in Section~\ref{sec:conclusion}.

\section{Research Setting}
\label{sec:research-method}
We briefly introduce our research questions and how we answered them. 
Then, we describe the dataset used in our research.

\subsection{Research Questions and Method}
Our work takes the first step toward extending issue duplication research to other link types in Issue Tracking Systems. 
We aim at understanding the various link types and how current duplicate prediction approaches can deal with them. 
In particular, we focus on the following research questions:  

\begin{itemize}
    \item \textbf{RQ1. Usage:} How are various link types used in practice, particularly in term of prevalence and structural properties? 
    \item \textbf{RQ2. Detection:} Are state-of-the-art models for duplicate detection able to distinguish between \dup{} and other links?
\end{itemize}

Studying the prevalence in RQ1 means analyzing how frequent certain link types are used in practice. 
By comparing the frequencies between different repositories, we hope to identify common trends. 
Moreover, structural properties refer to the structures of the different issue graphs that consist of certain link types. 
The idea here is that different ``graph types'' corresponding to different link types might exhibit different properties. 
These properties could be leveraged for prediction or to find problematic areas by investigating substructures that exhibit outlying values.


To answer RQ1, we first searched for a large ITS dataset, possibly of an issue tracker that allow customized link types and that is widely used in software development. 
The public JIRA dataset recently shared by Montgomery et al.~\citep{Montgomery:2022} meets this requirement. 
JIRA offers various default link types and allow their customization, e.g., to support different workflows. 
We manually analyzed all 75 link types in the dataset, their names, descriptions, and hundreds of random issue examples linked by the types. 
To reduce the comparison dimensions, we grouped link types that denote a similar connection to a link category. 
We set the categories based on the information a stakeholder would gain from viewing a link with a certain type as well as potential use cases of the link types. 
This step included multiple iterations between the first and second author and two main sessions with all authors to review and consolidate the categories and decide about five ambivalent types by discussing example links. 
The link types, categories, corresponding examples, and our notes are included in the replication package. 
We analyzed the general characteristics of the link categories. 
For this, we studied the issue graphs, where the vertices are the issues and the edges are the links that exist between these issues. 
We then compared the graphs spanned by a certain link type and a certain category.

RQ2 aims at re-evaluating state-of-the-art duplicate  prediction models by feeding them with other link types which are common in practice.
The link type \dup{} has been the focal point of research and has been rather viewed in isolation in previous work.
We aim to evaluate duplicate detection model's applicability in practice by using data that contains a more realistic distribution of link types.
Many duplicate link detection approaches use the Bugzilla datasets of Lazar et al.~\citep{Lazar:MSR:2014}.
These approaches are based on the assumption that \dup{} are more similar than any other pair of issues. 
The machine learning models in previous works are trained on datasets consisting of \dup{} and randomly created \NL{}, which are highly dissimilar in comparison.
Other link types are usually neither included in the training nor in the evaluation. 
We assume that linked pairs of issues should have a higher similarity than \NL{} and thus, can be harder to distinguish by the machine learning models.
Current research does not evaluate if the prediction  reliably predicts actual \dup{} or whether other link types in the ITS confuse the prediction.
Thus, we evaluate how well current duplicate prediction models work on our dataset which include issue pairs that are linked but are not duplicates.

\subsection{Research Data}

We used a dataset of 15 public JIRA repositories~\citep{Montgomery:2022}. 
Table~\ref{tab:DescStats} summarizes the analyzed repositories in terms of the number of issues, links, and link types. 
The table also shows the coverage, which represents the number of issues having at least one link as well as the share of cross-project links (within the same repository). 
The minimum and maximum for each metric across all repositories are indicated in bold.

The investigated repositories have different sizes ranging from 1,865 to 960,929 reported issues.
The repositories also vary in terms of link types and number of existing links: ranging from 4 link types and 44 links for Mindville up to 21 link types and 242,823 links in the case of Apache.
The average coverage is about~35\% meaning that only~35\% of all issues are linked. However, this varies too: Mindville has the lowest coverage of~4.0\% and Hyperledger 
the highest coverage of~55.5\%. 
The table shows that links rarely cross project boundaries as the majority (90\% average) of links are between issues of the same project, apart from Jira and RedHat.
The Jira repositories, corresponding to the development of the JIRA issue tracker itself, shows the second highest coverage among all repositories (55\%) as well as one of the highest number of link types~(19) after Apache~(21).

\begin{table}
\setlength\tabcolsep{2.5pt}  
\centering
\captionsetup{skip=6pt}
\caption{Overview of studied JIRA repos (alphabetical order).}\label{tab:DescStats}
\begin{tabular}{l rrrrr}
\multicolumn{6}{c}{\footnotesize{\makecell{Columns: Documented Link Types (\#Types);\\ percentage of issues with a link (\%Cov.);\\ percentage of links for issues from two different projects. (\%CP)}}}\\
\hline
\textbf{Repository} & \textbf{\#Issues} &  \textbf{\#Links} & \textbf{\#Types} &  \textbf{\%Cov.} & \textbf{\%CP}\\
\hline
 \rowcolor{hellgrau}Apache &  \textbf{970929} & \textbf{242823} &         \textbf{21} &            28.3\% &          5.3\% \\
  Hyperledger &   27914 &  16225 &          8 &            55.1\% &          4.6\% \\
    \rowcolor{hellgrau}IntelDAOS &    5557 &   3222 &         10 &            \textbf{55.5\%} &          - \\
        JFrog &   14769 &   3206 &         11 &            29.8\% &          8.2\% \\
         \rowcolor{hellgrau}Jira &  265343 &  98122 &         19 &            47.7\% &         \textbf{43.9\%} \\
JiraEcosystem &   40602 &  10911 &         18 &            32.8\% &          6.8\% \\
      \rowcolor{hellgrau}MariaDB &   31229 &  14618 &          8 &            44.5\% &          2.5\% \\
    Mindville &    \textbf{2134} &     \textbf{44} &          \textbf{4} &             \textbf{4.0\%} &          4.6\% \\
      \rowcolor{hellgrau}MongoDB &   90629 &  37545 &         13 &            42.6\% &         11.3\% \\
           Qt &  140237 &  35855 &          8 &            28.9\% &          7.2\% \\
       \rowcolor{hellgrau}RedHat &  315797 & 106200 &         18 &            38.9\% &         23.4\% \\
        Sakai &   49204 &  19057 &          7 &            42.2\% &          \textbf{1.4\%} \\
     \rowcolor{hellgrau}Sonatype &   77837 &   4289 &         11 &             7.5\% &          1.5\% \\
       Spring &   69100 &  14461 &          9 &            25.6\% &         10.0\% \\
   \rowcolor{hellgrau}SecondLife &    1865 &    630 &          6 &            39.8\% &          2.4\% \\
\hline
\textbf{Total} &  2103146  &  607208  & 171 &  - & -  \\
\textbf{Average} &  -  &  -  & 11.4 &  34.8\% & 9.5\%  \\
\hline
\end{tabular}
\end{table}

JIRA offers a general section for setting the links.
We encountered two additional link types that are bound to specific issue properties in our analysis: the \epi{} and \sub{} link types.
These types are not displayed in the general links section and require users to set the issue type  accordingly.
Apache, Qt, Secondlife, and Spring include rare cases where the \sub{} links are not linked to a subtask issue.

When mining the data for the links, we observed that a link can contain a private issue, for which we have no further information. 
We thus excluded from the analysis links where one of the linked issues is private.
Most linked issue pairs (97.6\%) had only one link type.
The remaining multi-links often seemed conflicting, such as \dup{} and \blo{}.
Thus we also removed them from the dataset for the subsequent analysis.

To simplify the analysis, we consider all connections as undirected.

\begin{table}
\centering
\small
\setlength\tabcolsep{1pt}  
\captionsetup{justification=centering, skip=6pt}  
\caption{Overview of cleaned link types, their descriptions, number of using projects, and the mapped  categories.}
\label{tab:LinkType}
\begin{tabularx}{\columnwidth}{l X c l}
\hline
\textbf{Link type} & \textbf{Description} & \textbf{Projects} & \textbf{Cat.}  \\ 
\hline
\rowcolor{hellgrau}Relate & Two issues are related to each other in a general way or stakeholders are unsure about the nature of the connection. & 15 & Rel \\
Duplicate & An issue is already in the system and someone added the same issue by accident. & 14 & Dup \\
\rowcolor{hellgrau}Subtask & An issue is a part of a larger issue/task. & 14 & Com \\
Clone & An issue was duplicated intentionally by the users, e.g., when an old bug resurfaces in a newer version. & 12 & Dup \\
\rowcolor{hellgrau}Depend & An issue cannot be resolved until another issue is resolved. & 10 & T/C \\
Epic & A link that consists of higher-level issue which has multiple low-level issues attached to it, similar to Parent-Child or Incorporate.  & 9 & Com \\
\rowcolor{hellgrau}Block & An issue cannot be resolved until another issue is resolved. & 8 & T/C \\
Incorporate & An issue is included or contained in another issue. E.g., an issue is used to collect multiple other issues, similar to Parent-Child. & 8 & Com \\
\rowcolor{hellgrau}Split & An issue is split into multiple issues to reduce the size or complexity. & 8 & Com \\
Cause & Issue A describes a problem that is at least a partial root cause for Issue B. & 7 & T/C \\
\rowcolor{hellgrau}Bonfire Testing & An issue was discovered while testing another issue. & 6 & WFl \\
Finish-finish & Two issues must be finished at the same time or patch. & 4 & T/C \\
\rowcolor{hellgrau}Supercede & An issue was refined based on existing issues.  & 4 & WFl \\
Finish-Start & An issue must be finished after another issue. & 3 & T/C \\
\rowcolor{hellgrau}Fix & An issue resolution fixes another issue as well. & 3 & WFl \\
Follow & An issue should be done after another issue, a temporal dependency. & 3 & T/C \\
\rowcolor{hellgrau}Parent-Child & A parent issue has multiple children issues. The children are parts of the parent.  & 3 & Com \\
Breaks & An issue breaks the solution of another issue.  & 2 & WFl \\
\rowcolor{hellgrau}Documented & An issue provides the documentation or log files for another issue or its fix.  & 2 & WFl \\
Implement & An issue implements another issue. & 2 & WFl \\
\rowcolor{hellgrau}Start-Start & Two issues must be started at the same time. & 2 & T/C \\
Test & An issue is testing the fix of another issue.  & 2 & WFl \\
\rowcolor{hellgrau}Backport & The same issue was fixed in an older version as part of the maintenance process. & 1 & WFl \\
Derive & An issue is derived from another issue.  & 1 & WFl \\
\rowcolor{hellgrau}Detail & An issue is detailed by another issue with more fine-granular description. & 1 & WFl \\
Precede & Issue A should be resolved before Issue B. & 1 & T/C \\
\rowcolor{hellgrau}Replace & An issue was created in order to replace an old one, often resolution is ``Duplicate''. & 1 & Dup \\
Require & An issue cannot be resolved until another issue is resolved.  & 1 & T/C \\
\rowcolor{hellgrau}Start-Finish & The completion of an issue sets off the start of another. & 1 & T/C \\
Trigger & While working on an issue, another issue was triggered. & 1 & WFl \\
\hline
\end{tabularx}
\end{table}

\section{Usage of Issue Links Types (RQ1)}\label{sec:link-types}
We first present the various link types, the categories emerging from the manual analysis as well as the corresponding usage frequencies. Then, we report on the structural properties observed for the types.  

\subsection{Link Categories and their Prevalence}

We recorded a total of 75 different link type names in the original data. 
We first manually grouped together link types with similar word stems (e.g. ``finish-finish [gantt]'', ``gantt: finish-finish'' or ``Depend'', ``Dependency'', ``Dependent'', ``Depends''). 
Then, we compared the meaning of the types by reviewing their textual descriptions in the repositories and checking random examples of linked issues. 
For instance, 9 repositories used synonym names for \inc{}, such as ``includes'',  ``contains'' or ``incorporates''.
This step led to 30 distinct link types, which are listed in 
Table~\ref{tab:LinkType} with brief descriptions, and the count of repositories where they are used.

\begin{table*}[ht]
\setlength\tabcolsep{2.5pt}  
\centering
    \caption{Popularity of the link types across the studied JIRA repositories (usage shares in \%).}
    \label{tab:LtProjectsShare}
    \begin{tabularx}{\textwidth}{l rrrrrrrrrr r X}
    \hline
\textbf{Repository}& \textbf{Relate} & \textbf{Subtask} & \textbf{Duplicate} & \textbf{Clone} &  \textbf{Depend} & \textbf{Epic} &  \textbf{Split}  &  \textbf{Block} & \textbf{Incorporate} & \textbf{Cause} & \textbf{Coverage} & \textbf{Other types}\\
         \hline
\rowcolor{hellgrau}Apache        &  28.5\% &   32.5\% &    10.2\% &   1.7\% &   5.2\% &   4.9\% &   - &   6.2\% &       4.1\% &  1.0\% &   94.4\% & -\\
Hyperledger   &  17.1\% &  27.6\% &     3.9\% &    2.9\% &    - &  39.7\% &  0.5\% &   8.3\% &        - &   - &   99.9\% & - \\
\rowcolor{hellgrau}IntelDAOS     &  12.1\% &    5.8\% &     3.6\% &    4.5\% &    - &  65.7\% &   - &   6.6\% &        - &   - &   98.4\% & -\\
JFrog         &  27.2\% &  36.3\% &    19.9\% &   0.6\% &   8.0\% &    - &   - &    - &       1.3\% &   - &   93.4\% & \footnotesize Trigger\\
\rowcolor{hellgrau}Jira          &  64.0\% &   2.5\% &    21.8\% &   2.8\% &   0.1\% &    - &  0.2\% &   1.0\% &       2.5\% &  1.8\% &   96.4\% & -\\
JiraEcosystem &  23.3\% &  20.4\% &    15.8\% &   1.8\% &   1.1\% &  23.3\% &   1.2\% &   5.9\% &       1.8\% &  3.8\% &   98.2\% & - \\
\rowcolor{hellgrau}MariaDB       &  51.1\% &    6.1\% &      9.4\% &    - &    - &   6.4\% &  0.2\% &  13.0\% &       7.9\% &  6.0\% &   100.0\% & - \\
Mindville     &  43.2\% &    - &    38.6\% &  15.9\% &    - &    - &   - &   2.3\% &        - &   - &   100.0\% & -\\
\rowcolor{hellgrau}MongoDB       &  49.1\% &   1.6\% &    17.4\% &   0.2\% &  18.5\% &   6.3\% &  0.4\% &    - &        - &  1.7\% &   95.2\% & -\\
Qt            &  22.4\% &   23.8\% &    10.7\% &    - &  16.4\% &  13.2\% &  6.4\% &    - &        - &   - &   92.8\% & \footnotesize Replace\\
\rowcolor{hellgrau}RedHat        &   26.6\% &  20.4\% &     5.1\% &  15.6\% &    - &    - &   0.1\% &  15.3\% &       9.1\% &  2.8\% &   95.0\% & \footnotesize Follow\\
Sakai         &  48.7\% &  17.6\% &     9.5\% &   4.8\% &  12.9\% &    - &   - &    - &       6.7\% &   - &   99.99\% & -\\
\rowcolor{hellgrau}SecondLife    &  29.4\% &  49.8\% &      - &   7.6\% &   4.4\% &    - &   - &    - &       2.2\% &   - &   93.5\% & \footnotesize Parent-Child\\
Sonatype      &   38.8\% &  31.0\% &     7.9\% &    - &   3.7\% &   0.2\% &  0.1\% &    - &        - &  5.3\% &   86.9\% & \footnotesize Bonfire Testing, Fix, Supercede\\
\rowcolor{hellgrau}Spring        &  47.7\% &  13.4\% &    12.1\% &    0.1\% &  12.1\% &  11.3\% &   - &    - &        - &   - &    96.7\% & \footnotesize Supercede\\
\hline
\textbf{Mean}           & 35.3\%   & 20.6\%  & 13.3\% & 4.9\% & 8.2\% & 19.0\% & 1.1\% & 7.3\% & 4.5\% & 3.2\% & 96.1\% & - \\
\textbf{Stand. Dev.}    & 14.8\%   & 14.1\%  & 9.2\% & 5.5\% & 6.4\% & 21.2\% & 2.2\% & 4.9\% & 3.0\% & 1.9\% & 3.6\% & - \\
\hline
    \end{tabularx}
\end{table*}


Unsurprisingly, link types provided by default in the JIRA ITS are among the most popular. 
Moreover, the custom link types \dep{}, \inc{}, \spl{}, and \cau{} are common. 
Table~\ref{tab:LtProjectsShare} shows the shares of the common link types, appearing in 7 or more repositories. 
Our analysis shows that \rel{} is overall predominant with~35\% of all links. 
In five repositories, namely Jira, MariaDB, MongoDB, Sakai, and Spring even~>47\% of the links are \rel{}. 
On average, \sub{} has a share of~21\%, \epi{}~19\%, and \dup{}~13\% of all links. All other types have a share of less than~10\%.
We also observe that \dep{} and \blo{ } are complementary in most cases. The only ITS that uses both with a similar frequency is Apache. 
The same pattern is also visible in \inc{} and \spl{} except for JiraEcosystem.

After a close inspection of multiple examples for each of the 30 cleaned link types in the corresponding ITS, 
we identified the following five overarching link categories:  
\begin{itemize}
    \item[\textbf{\Rel{} (Rel)}:] Links that indicate a general relationship between issues, which is not specific, but offers relevant or useful information to implement, fix, reproduce, or plan an issue. E.g., one java class is the origin of two bugs.
    \item[\textbf{\Dup{} (Dup)}:] Links to indicate that two issues describe the same bug, feature, or task. A duplication issue can be added intentionally (Clone, Replace) or unintentionally (Duplicate). E.g., two users find and report the same bug without being aware that the bug was already documented in the ITS.
    \item[\textbf{\Tem{} (T/C)}:] Indicate that an issue depends on the resolution of another. This can be due to causal or temporal reasons. E.g., One feature is part of a release. A second feature that builds on the first can only be released later.
    \item[\textbf{\Com{} (Com)}:] Links that indicate a structural relationship between two issues. This link is created when an issue is split into multiple sub-issues, when user stories are added to an epic, or a parent issue gets split to sub-tasks or children issues. E.g., a stakeholder creates a ticket for a feature and splits it into multiple tasks that need to be completed.
    \item[\textbf{\Wor{} (WFl)}:] Links that indicate that an issue was discovered, reported, fixed, refined, or documented while working on a certain development activity on another issue, like testing, reviewing, or implementing it. E.g.,~while testing a feature, a developer finds an unknown bug and documents it.
\end{itemize}

Table~\ref{tab:LinkType} shows the mapping of the link types to the categories. 
A few link types might seem harder to categorize as the names show a potential overlap with another category. 
For instance, we categorized ``Supercede'' as \Wor{} as it reports the refinement or evolution of an issue over time and is created while working on the issue. 
Similarly, the link type ``Break'' can be misleading at first, since it does not denote a work breakdown but a regression. 
For such cases, we determined the final categorization based how the example links were actually used in the studied repositories.
An interesting case are \clo{} and \dup{} types. 
On the first glance they seem to describe the same type, but on closer inspection we found that \clo{} links are created automatically, when an issue is cloned with the corresponding JIRA feature.
A stakeholder creates the same issue again and changes either the properties or the title based on the issue to be cloned.
In contrast, \dup{} links are created unintentionally when two different stakeholders submit the same issue without being aware of the other.
Both \clo{} and \dup{} are thus part of \Dup{}, but they should not be merged as classes in a more fine-grained link type detection system.

\begin{figure}
        \includegraphics[width=1.05\columnwidth]{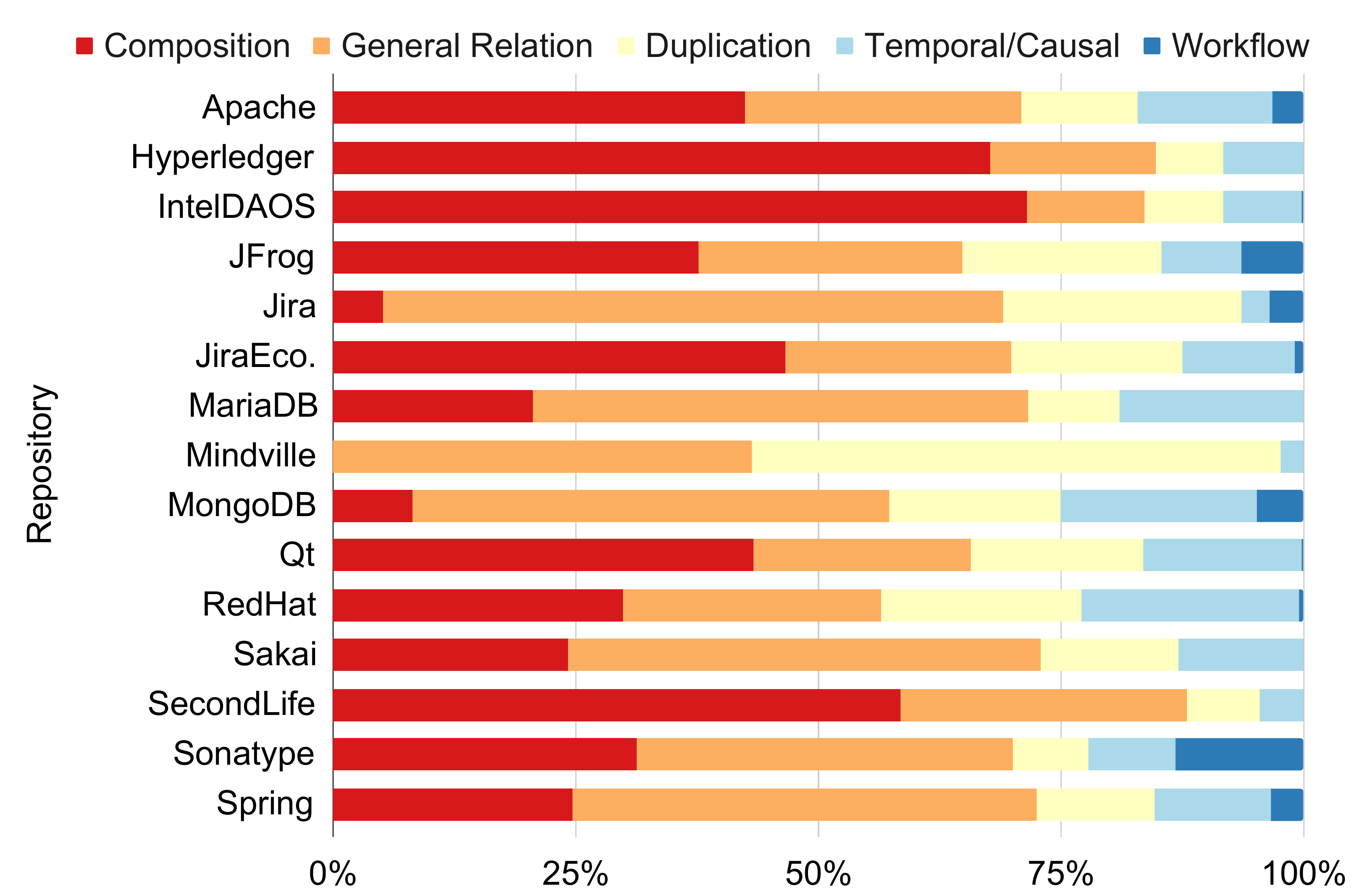}
       \caption{Share of link categories across  studied repositories.}\label{fig:LinkComp}
       \vspace{-0.7cm}
\end{figure}

Figure~\ref{fig:LinkComp} depicts the prevalence of the link categories in each repository.
All categories are used with different frequencies  across all repositories, except that 
 \Com{} was not used by Mindville and \Wor{} was not used by MariaDB, Mindville, and SecondLife.
Furthermore, we observe three popularity levels. First, \Com{} and \Rel{} were fairly pervasive: on average 36.6\% of all links were \Com{} closely followed by \Rel{} which accounts for 35.3\% of all links.
Then, \Dup{} and \Tem{} account on average for 16.8\% and 11.4\% of all links in the repositories respectively. 
\Dup{} has one outlier, Mindville, which only has 46 links. 
Finally, \Wor{} is least used link category averaging about 3.0\% of all links. 
\Wor{} has one outlier, Sontatype, which uses this category for about 13.1\% of the links. 

\vspace{6pt}
\textbf{In summary}, the results confirms the heterogeneity of link types in practice and that \dup{} links are by far not the most frequent links. 
The fact that there are, on average, $\sim$11 link types per repository suggests either an overwhelming semantic complexity, an unnecessarily bloated complexity with potential type mismatch, or ineffective issue linking.

\subsection{Comparison of Link Types Structures}\label{sec:link-graphs}
Next we explore and compare the graph structures formed by the different link categories. 
For this, we use metrics which are fairly popular and describe basic graph properties in the field of graph theory. 
We chose these metrics as they can tell us how the issue graphs generally look like without visualizing them completely.
First \textbf{Complexity} describes how large the  graph components are, revealed by four metrics:
    \begin{itemize}
        \item \textbf{\%Isolated:} The share of issues without links. A high number of isolated issues shows that most issues are without links.
        \item \textbf{\%2Comp:} The amount of graph components with exactly two or more issues. A high number of 2-issues-components shows that the graph is fairly simple.
        \item \textbf{\%3Comp+:} The amount of components in the graph with exactly three or more issues. A high number of 3-or-more-issue-components shows that the graph is rather complex.
        \item \textbf{Average Density:} The average number of links in a component divided by the number of possible links (only for components with three or more issues). A high value indicates that the graph components are highly inter-connected.
    \end{itemize}
    
\begin{figure}[b]
\centering
\captionsetup[subfigure]{justification=centering}
\begin{subfigure}[b]{0.21\textwidth}
    \includegraphics[scale=0.21]{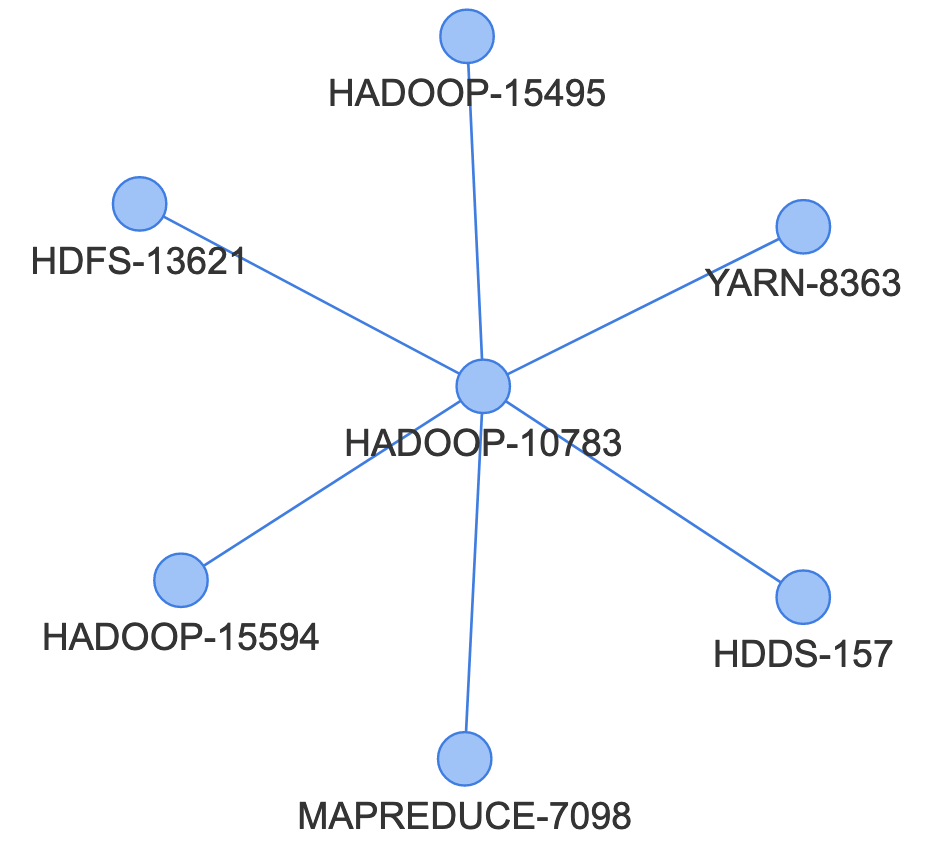}
    \caption{Star graph of \Com{} category in Apache.}
    \label{fig:star_incorporate}
\end{subfigure}
\hspace{1em}
\begin{subfigure}[b]{0.2\textwidth}
    \centering
    \includegraphics[scale=0.24]{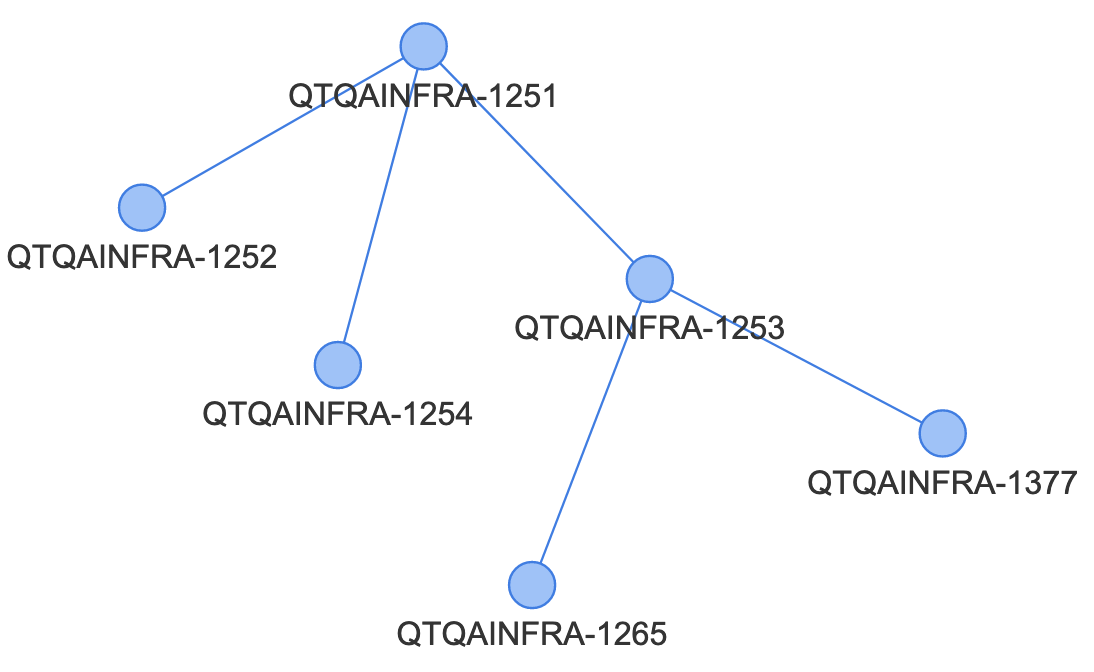}
    \caption{Tree graph of \Tem{} category in Qt.}
    \label{fig:irregular_duplicate}
\end{subfigure}
\caption{Example graph structures for different link categories.}
\label{fig:perception_suggestion_types}
\vspace{-0.5cm}
\end{figure}

The graph \textbf{Shape} is, characterized by two metrics:
    \begin{itemize}
        \item \textbf{\%Trees:} Number of tree-graph components. If the graph components mostly consist of tree graphs, the issues are ordered hierarchically.
        \item \textbf{\%Stars:} Number of star-graph components, a subtype of trees. A high percentage of stars indicate central issues, connected to many other issues which do not have connections to each other.
    \end{itemize}
Figure~\ref{fig:perception_suggestion_types} shows example structures found in the issue graphs: star shape for \Com{} links in Apache and a tree shape for \Tem{} links in Qt.
We analyze only the connected components. 
As isolated issues and 2-issues-components are simple structures, we exclude them in the shape analysis and focus on larger structures including 3 or more components.

\textbf{Assortativity} is the Pearson correlation between the degree of nodes at both ends of each edge in the graph. This value is between $-1$ and $1$. While social networks are usually more assortative (positive values), technological and biological networks tend to be more disassortative (negative values). Assortative networks are more robust to vertex removal. Disassortative networks often have strong hierarchical configurations with large nodes connecting smaller nodes~\citep{Newman:APS:2002}.

Finally, \textbf{Transitivity} is  3 times the number of triangles divided by number of triads (two links that share one issue)~\citep{Schank:2005}. This value is between $0$ and $1$. Transitivity measures the frequency of triangles. A value near $0$ means the graph has low transitivity and a value near $1$ means the graph has high transitivity. E.g., if a link type has high transitivity, then if issue A is linked with issue B which is in turn linked to issue C, that stakeholders also put the same link type between A and C in the ITS.

\begin{table}[ht!]
 \small
\setlength\tabcolsep{2.5pt}  
\centering
\captionsetup{skip=6pt}
    \caption{Metrics for the entire issue graph in a repository. All values in percent except assortativity.}\label{tab:overviewWholeGraph}
\begin{tabularx}{\columnwidth}{X | r r r r | r r | r }
\hline
    \multicolumn{8}{c}{\footnotesize{\makecell{Columns: Assortativity (Asso.); Average Density (Den.); Isolated (\%Iso.);\\ 2-Components (\%2C); 3-or-more-Components (\%3C+)}}}\\
    \hline
\multirow{2}{*}{\textbf{Repo.}} & \multicolumn{4}{c}{\textbf{Complexity}}&\multicolumn{2}{|c|}{\textbf{3C+ Shape}}&\multirow{2}{*}{\textbf{Asso.}}\\
\cline{2-7}
&\%Iso.&\%2C&\%3mC & Den. &\% Tree&\% Star&\\

\hline
\rowcolor{hellgrau}Apache          &       71.7          &      63.0  &      37.0  &  .500&     82.5  &    61.8  &  -.061\\
Hyperledger     &       44.9           &      47.3  &      52.7  &  .428&     78.4  &    62.2  &  -.112\\
\rowcolor{hellgrau} IntelDAOS      &       \textbf{44.5 } & \textbf{39.9 } & \textbf{60.1 } &\textbf{.374} &75.7  &    62.9  &-.233\\
 JFrog           &       70.2           & 66.2  & 33.8  & .556 & \textbf{87.1 } &  \textbf{66.3  } & -.148\\
\rowcolor{hellgrau} Jira            & 52.3                 &\textbf{78.7 } & \textbf{21.3 } & .534 & \textbf{65.2 } & \textbf{47.6 } & .003\\
 JiraEco.   &       67.2           &      59.6  &      40.4  &   .495 & 83.0  &    63.4   &      -.148\\
\rowcolor{hellgrau} MariaDB         &       55.5           &      60.3  &      39.7  &     .543 &   75.1  &    57.1  &  -.061\\
 Mindville       &  96.0                & 95.1  &       4.9  &   \textbf{.583} &    100.0  &   100.0  &  -.055\\
\rowcolor{hellgrau} MongoDB         &  57.4                & 61.8  &      38.2  &    .524 &      80.0  &    54.8  & \textbf{.005}\\
 Qt              &       71.1           &      63.8  &      36.2  &     .522 &    85.6  &    65.8  & -.157\\
\rowcolor{hellgrau} RedHat          &       61.1           &      61.8  &      38.2  &    .523 &   78.4  &    58.0  &   -.054\\
 Sakai           &       57.8           &      60.7  &      39.3  &    .514 &  74.4  &    52.5  &    -.079\\
\rowcolor{hellgrau} SecondLife      &  60.2                & 52.8  &      47.2  &      .450 &   77.9  &    66.2   & \textbf{-.282}\\
 Sonatype        &  \textbf{92.5 }      & 68.1  &      31.9  &    .502 &   84.2  &    61.8   &   -.129\\
\rowcolor{hellgrau} Spring          &       74.4           &      62.6  &      37.4  &     .520 &    81.2  &    62.6   & -.099\\
\hline
\textbf{Mean} &       65.1             &      62.8  &      37.2  &     .505 &  80.6  &    62.9  &   -.107\\
\hline
\textbf{Std. Dev.} &  15.0          &      12.6  &      12.6  &      .052 &   7.6  &    11.6   &   .079\\
\hline

\end{tabularx}
\end{table}

We first analyze the structure of the whole repositories based on all links (without distinguishing between categories). 
Table~\ref{tab:overviewWholeGraph} shows the metrics for the complete issue graph (except for transitivity which is rather meaningless without referring to the concrete link semantic). 
We highlight the min and max values for each metric and also present the mean and standard deviation. Overall the values seem rather homogeneous only with a few exceptions.  
Most repositories have a high number of isolated issues, which shows overall a very disconnected repository structure. 
Furthermore, the repositories seem to share almost a 2/3-1/3 split concerning their simple 2-issue components to the more complex components consisting of at least~3 issues.
The outliers here are Jira, which seems to have a relatively high number of 2-issue components -- despite having the second highest number of coverage. 
Hyperledger and IntelDAOS both have a low number of simple components and boast more complex structures. 

Assortativity is around -0.1 on average.  
SecondLife is the most dissortative issue graph (-0.282) which could indicate a very hierarchical structure whereas MongoDB has the highest assortativty (0.005). 
It seems that ITS repositories tend towards hierarchical structures.
Additionally, the average density of the three or more issue components is around 0.505, which means that about half of the theoretically possible edges are indeed set as issue links. 
On average, 80.6\% of the more complex components are trees, which indicates that transitive edges might not be explicitly contained in the studied ITS.
Furthermore around 62.9\% of these components are star shaped. 

\begin{table}[b]
\small
\setlength\tabcolsep{2.0pt}  
\centering
\captionsetup{skip=6pt}
    \caption{Metrics comparison of link categories.}\label{tab:CatMetricsComparison}
\begin{tabularx}{\columnwidth}{X|rrrrr|rr|rr}
    \hline
    \multicolumn{9}{c}{\footnotesize{\makecell{Columns: Assortativity (Asso.); Transitivity (Tran.);\\ Average Density(Den.); Isolated (\%Iso.);\\ 2-Components (\%2C); 3-or-more-Components (\%3C+)}}}\\
    \hline
\multirow{2}{*}{\textbf{Category}} & \multicolumn{5}{c}{\textbf{Complexity}}& \multicolumn{2}{|c|}{\textbf{3C+ Shape}}&\multirow{2}{*}{\textbf{Asso.}}&\multirow{2}{*}{\textbf{Tran.}}\\
\cline{2-8}
                &\%Iso.    &\%2C  &\%3C+  &Den. & \#Links &\%Tree&\%Star&& \\
         \hline
\rowcolor{hellgrau}\textit{Rel.}&       85.6           & 70.7           & 29.3            & \textbf{.562}             & 218833            & \textbf{81.9}            & \textbf{62.8}            & \textbf{.025} & \textbf{.121}    \\
\textit{Dup.}&       92.6           & 82.8   & 17.2   & .558             & 98178             & 88.5            & 76.4            & .003 & .050    \\
\rowcolor{hellgrau}\textit{T/C.}&       94.7          & 73.5            & 26.5            & .496             & 84236             & 84.4            & 63.7            & -.049 & .048   \\
\textit{Com.}&       85.0           & \textbf{41.2}   & \textbf{58.8}   & \textbf{.394}    & 190913            & \textbf{97.7}   & \textbf{90.1}   & \textbf{-.260} & \textbf{.002}   \\
\rowcolor{hellgrau}\textit{WFl}&       98.9           & \textbf{84.5}            & \textbf{15.5}            & .490             & 15048             & 82.5            & 75.6            & -.098 & .003    \\
\hline
    \end{tabularx}
\end{table}

In the next step, we calculated and compared the metrics for each subgraph spanned by one link category.
Table~\ref{tab:CatMetricsComparison} shows the results.
We observe that \Rel{} differs from the other categories with the highest transitivity score. This is rather surprising since we expected ``stronger'' links such as temporal, causal or composition links to be more transitive than general relation. 
\Dup{}s tend to be simpler, only having 17.5\% complex components. 
These complex components are often star-shaped (76.4\%), which means there exists one original issue, which is connected to its duplicates, which by themselves do not share a link. 
As expected, \Com{} have the most tree-like structures with 90.1\% of their components being cycle-free. 
This also means 9.9\% of the complex issue composition components contain a cycle, which is rather surprising given the meaning of these links and might reveal wrong link usages. 
\Com{} also has the lowest assortativity with {-0.260}, which indicates that this category is very hierarchical in nature compared to the others. 
\Tem{} links are less star-shaped than other categories, which suggests some "string-like" substructures as the link category is rather tree-like.
\Wor{}'s metrics are in between the other metrics.

We observe that the percentage of isolated issues is negatively correlated with the number of links. 
A high number of isolated issues coupled with a high number of links reveals that the connected components of the graph are very clustered.
From the lower isolated issues and higher amount of links, we conclude that these links are rather scattered. 
The 0.5 average density confirms this.

\vspace{6pt}
\textbf{In summary}, the repositories are overall structurally similar,  
while the link categories differ from each other.
The differences seem not strong enough to inform a link type prediction. 
These metrics also give us a starting point for baselines that represent the structural and complexity characteristics of ITS.

\section{Link vs Link Type Prediction (RQ2)}
\label{sec:link-prediction}

Due to the considerable variety of link types in JIRA, it is important that duplicate detection approaches (representing a well-researched and mature area) are also able to differentiate \Dup{} links from other link types. 
Thus, this section aims to explore how well current duplicate detection models perform on our dataset. 

\subsection{Experimental Setting}

\subsubsection{Models}
We used the Dual-Channel CNN (DCCNN) approach by He et al.~\citep{He:ICPC:2020} 
 and a Single-Channel CNN (SCCNN) approach following recent work by Deshmukh et al.~\citep{Deshmukh:ICSME:2017} and Budhiraja et al.~\citep{Budhiraja:ICSE:2018} .
 A DCCNN takes two issues and embeds them together, thus modeling a link in the embedding. 
 In contrast, a SCCNN only embeds one issue and then takes two issue embeddings and uses them to predict if these issues form a duplicate link.
To the best of our knowledge, the Dual-Channel approach by He et al.~has the highest  performance in the duplicate prediction literature. 
All other recent top-performing models use a Single-Channel approach tuned with varying layers and CNN or LSTM layers~\citep{Deshmukh:ICSME:2017}\citep{Budhiraja:ICSE:2018}.

All published DCCNN and SCCNN models, which we are aware of,  use the same training and evaluation set-up: use duplicates instance for the positive class and randomly created non-links for the negative class.
To evaluate current models on our dataset, we asked the authors for the source code. 
We were only able to receive the original code of the DCCNN approach from one of the authors, which we used in our experiment. 
For the SCCNN, the researchers could not share their original code.
Thus, we implemented a SCCNN architecture with the same layers as the DCCNN model. This bears the advantage that the results between both approaches are comparable and we can isolate the effect of evaluating a more realistic dataset with various link types, which is a common case in practice.

We used the title and description of the issues as input to predict if these are connected via \Dup{} links or not.
The evaluated DCCNN model stacks both issues of a link resulting in a tensor of dimensions (300, 20, 2) and uses this as inputs for the CNN layers.
The SCCNN model has a (300, 20, 1) input. It uses the same CNN layered architecture as the DCCNN model. 
Both issues are processed by the model and then concatenated and fed into a dense neural network.

For the data preprocessing we followed the same steps as He et al.~\citep{He:ICPC:2020} to generate comparable results.
We also used the same hyper-parameter configurations as He et al., which are also recommended by Zhang and Wallace~\citep{Zhang:IJCNLP:2017}.
We removed stopwords, lowercased, and stemmed the textual data from description and title to train a Word2Vec model.
This is usually done for all state-of-the-art models.
Currently, BERT models achieve higher performance and are favored for these tasks, but since the original research uses Word2Vec, we chose to also use Word2Vec.

We carried out these steps for each repository.
The initial dataset crawled from the JIRA repositories contain \Dup{} links and \OL{}.
We also generated \NL{}, by randomly choosing from all closed issues without the resolution ``Duplicate''.
Our replication package include all source code details, data, and configuration files.

\subsubsection{Data}
We trained the two models on 14 repositories. 
We excluded Mindville, as it only has 46 links in total. Additionally, this repository has only one data point with the type \blo{}. 
Technically, the models could be able to learn on this small dataset.
However, the test set only contains 9 data points, which is too small to draw any significant conclusions.
Furthermore, if a model only predicted one class for all samples in the test data or had an infinite loss, we labeled it as unable to learn on the training data.
We excluded the repository from this model's result in this case.
For the SCCNN model this was only SecondLife.
For the DCCNN model these were IntelDAOS, JFrog, SecondLife, and Sonatype: all repositories with less than 5,000 links. 
Essentially, we distinguish between 3 different classes: \Dup{}, \OL{} (links of a different type), and \NL{}, while state of the art duplicate detection models are trained on \Dup{} vs. \NL{} and the \OL{} are disregarded.

We split the data once randomly in an 80/20 training-test-split.
The test set is balanced in the three classes \Dup{}, \OL{}, and \NL{}.
We report the results on this ``new'' test set, which includes \OL{} as well as the ``traditional'' test set, which only consists of \Dup{} and \NL{}.
This way we can evaluate how the models will perform on a realistic  dataset.

\subsubsection{Training Sets} For the training set we used 3 different configurations to analyze what learning task the models perform best on.
Specifically, we wanted to find out if the models are able to differentiate \Dup{} from other link types or if they simply learn if two issues are linked. 
The training set has always two labels. 

We experimented with three different training set configurations. 
The first two training sets aim at distinguishing \Dup{} from [\OL{} and \NL{}], with \Dup{} having label 1 and the other two classes having label 0. 
The third training set aims at distinguishing links from the \NL{}. Thus \Dup{} and \OL{} have label 1 and \NL{} has label 0. 
For reporting the results we used the abbreviations explained in the following:
\begin{itemize}
    \item[\textbf{DvsNL}] \Dup{} vs. \NL{}. 
    This correspond to the common evaluation setting from the literature (e.g.~\citep{He:ICPC:2020},\citep{Deshmukh:ICSME:2017},\citep{Budhiraja:ICSE:2018}).
    We check if the models are robust and still able to correctly classify \OL{} and how the performance changes in comparison to the ``traditional'' test set.
    \item[\textbf{DvsOLNL}] \Dup{} vs. [\OL{} \& \NL{}].
    We use \Dup{} links as one class and \NL{} and \OL{} together as the second class.
    We check if the models are able to differentiate \Dup{} from \NL{} and \OL{} when trained on all classes.
    \item[\textbf{DOLvsNL}] [\Dup{} \& \OL{}] vs. \NL{}.
    We check how well the models perform when tasked with distinguishing linked issues from non-linked issues.
\end{itemize}
Each training set is balanced including approximately the same amount of instances for each class. 
This means we did not include all possible \NL{} and \OL{}, as this would heavily imbalance the dataset and possible skew the performance.

\subsubsection{Metrics}
We report on the precision, recall, and F1-score of each class, and on the accuracy
for the SCCNN and DCCNN models. 
We also calculated the macro averages for precision, recall, and F1-score of the classes as we do not want to give one class more weight than the others.
We did this for both test sets and also report the changes of the model performance from the ``traditional'' to the ``new'' set.

\subsection{Experiment Results}

Table~\ref{tab:DvsNLResults} shows the results when trained on the common settings from the literature, i.e.~the models learn from a training set consisting of \Dup{} and \NL{}.
We observe, as expected, that the performance on the traditional test set is high, with a Macro F1-score of 0.80 (SCCNN) and 0.74 (DCCNN).
Once the models were presented with \OL{} data, the performance drops with 9\% lower F1-score for the SCCNN and 7\% for the DCCNN.

\begin{table}[h]
\small
\setlength\tabcolsep{2.5pt}  
\centering
\captionsetup{skip=6pt}
\caption{Single Channel CNN and Dual Channel CNN model results averaged over the repositories for the case DvsNL.}\label{tab:DvsNLResults}
\begin{tabularx}{\columnwidth}{l l L L L L} 
\hline
& \textbf{Class} & \textbf{Pre.} & \textbf{Rec.} & \textbf{F1} & \textbf{Acc.}\\
  \hline
  \multicolumn{6}{c}{On traditional test set only with \Dup{} and \NL{}}\\
  \hline
 \parbox[t]{2mm}{\multirow{3}{*}{\rotatebox[origin=c]{90}{\footnotesize SCCNN}}} &
\cellcolor{hellgrau}\Dup{} & \cellcolor{hellgrau}.78 & \cellcolor{hellgrau}.85 & \cellcolor{hellgrau}.81 & 
 \multirow{3}{*}{.80}\\ 
 & \NL{} & .83 & .75 & .78 & \\
\cline{2-5}
 & \cellcolor{hellgrau}Macro avg & \cellcolor{hellgrau}.81 & \cellcolor{hellgrau}.80 & \cellcolor{hellgrau}\textbf{.80} & \\
 \hline
 \parbox[t]{2mm}{\multirow{3}{*}{\rotatebox[origin=c]{90}{\footnotesize DCCNN}}} & 
 \Dup{} & .76 &  .70 & .72 & 
 \multirow{3}{*}{.74}  \\ 
 & \cellcolor{hellgrau}\NL{} &\cellcolor{hellgrau} .72 & \cellcolor{hellgrau}.78 & \cellcolor{hellgrau}.75 &  \\
\cline{2-5}
 & Macro avg & .74 & .74 & \textbf{.74}  & \\
  \hline
  
 \multicolumn{6}{c}{On new test set including \OL{}}\\
 \hline
 \parbox[t]{2mm}{\multirow{3}{*}{\rotatebox[origin=c]{90}{\footnotesize SCCNN}}} &
 \cellcolor{hellgrau}\Dup{} & \cellcolor{hellgrau}.52 (-.26) & \cellcolor{hellgrau}.85 (+/-0) & \cellcolor{hellgrau}.64 (-.17) & 
 \multirow{3}{*}{.68 (-.12)} \\ 
 & \textit{Non-Dup.}& .89 (+.06) &  .59 (-.16) & .70 (-.08)& \\
\cline{2-5}
 & \cellcolor{hellgrau}Macro avg & \cellcolor{hellgrau}.70 (-.11) & \cellcolor{hellgrau}.72 (-.08) & \cellcolor{hellgrau}\textbf{.71 (-.09)} & \\
 \hline
 \parbox[t]{2mm}{\multirow{3}{*}{\rotatebox[origin=c]{90}{\footnotesize DCCNN}}} & 
 \Dup{} & .51 (-.25) &  .70 (+/-0) & .59 (-.13) & 
 \multirow{3}{*}{.68 (-.06)}  \\ 
 & \cellcolor{hellgrau}\textit{Non-Dup.}  & \cellcolor{hellgrau}.82 (+.10) & \cellcolor{hellgrau}.67 (-.11) & \cellcolor{hellgrau}.73 (-.02) & \\
\cline{2-5}
 & Macro avg & .67 (-.07) & .68 (-.06) & \textbf{.67 (-.07)}  & \\
 \hline
\end{tabularx}
\end{table}

On average, 57.5\% (SCCNN) and 44.2\% (DCCNN) of the \OL{} were classified as \Dup{}.
We see that the DCCNN architecture is slightly better at distinguishing \Dup{} from \NL{}, but neither architecture is particularly good at this.
We also observe that the model performance regarding \OL{} varies a lot between the repositories, ranging from 17\% up to 85\% (SCCNN) and 32\% up to 77\% (DCCNN).

There were also differences depending on the link categories.
\Com{} links were mistaken as \Dup{} in 34\% of cases, \Rel{} in 52\% of cases, \Tem{} in 40\%, \Wor{} in 45\% for the DCCNN model. For the SCCNN model, \Com{} links were mistaken as \Dup{} in 40\% of cases, \Rel{} in 47\% of cases, \Tem{} in 49\%, \Wor{} in 49\%.
The hypothesis underlying current approaches is that \Dup{} text descriptions are highly similar while \NL{} which are highly dissimilar. The prediction models work thus well when only evaluated on these two classes -- even if not as precise as for the Bugzilla dataset.
It seems that \OL{} have varying degrees of similarity across the different repositories.

Overall, the models are not able to distinguish \OL{} from \Dup{}.
This suggests we should include the \OL{} in the training set to enable the models to differentiate them from \Dup{}.
For this we have two options, 1) train the model to differentiate \Dup{} from \OL{} and \NL{} and 2) train the models to differentiate \Dup{} and \OL{} (linked) from \NL{}.

Table~\ref{tab:DvsNLOLResults} presents the results of the first option, while Table~\ref{tab:DOLvsNLResults} reports on the second.
We observe in Table~\ref{tab:DvsNLOLResults} that the model performs worse on the traditional test set compared to Table \ref{tab:DvsNLResults}, but becomes better on the new test set.
Overall, on this training set configuration DvNLOL the performance also worsens slightly on the new test data, decreasing only 0.02 (SCCNN) and 0.01 (DCCNN).

\begin{table}[b]
\small
\setlength\tabcolsep{2.5pt}  
\centering
\captionsetup{skip=6pt}
\caption{Single Channel CNN and Dual Channel CNN model results averaged over the repositories for the case DvsNLOL.}\label{tab:DvsNLOLResults}
\begin{tabularx}{\columnwidth}{l l L L L L} 
\hline
& \textbf{Class} & \textbf{Pre.} & \textbf{Rec.} & \textbf{F1} & \textbf{Acc.}\\
  \hline
  \multicolumn{6}{c}{On traditional test set only with \Dup{} and \NL{} }\\
  \hline
 \parbox[t]{2mm}{\multirow{3}{*}{\rotatebox[origin=c]{90}{\footnotesize SCCNN}}} &
 \cellcolor{hellgrau}\Dup{} & \cellcolor{hellgrau}.73 & \cellcolor{hellgrau}.81 & \cellcolor{hellgrau}.77 & 
 \multirow{3}{*}{.75}\\ 
 & \NL{} & .78 &  .68 & .72 & \\
\cline{2-5}
 & \cellcolor{hellgrau}Macro avg & \cellcolor{hellgrau}.76 & \cellcolor{hellgrau}.75 & \cellcolor{hellgrau}\textbf{.75} & \\
 \hline
 \parbox[t]{2mm}{\multirow{3}{*}{\rotatebox[origin=c]{90}{\footnotesize DCCNN}}} & 
 \Dup{} & .70 &  .70 & .70 & 
 \multirow{3}{*}{.70}  \\ 
 & \cellcolor{hellgrau}\NL{} & \cellcolor{hellgrau}.70 & \cellcolor{hellgrau}.70 & \cellcolor{hellgrau}.69 &  \\
\cline{2-5}
 & Macro avg & .70 & .70 & \textbf{.70 } & \\
  \hline
  
 \multicolumn{6}{c}{On new test set including \OL{}}\\
 \hline
 \parbox[t]{2mm}{\multirow{3}{*}{\rotatebox[origin=c]{90}{\footnotesize SCCNN}}} &
 \cellcolor{hellgrau}\Dup{} & \cellcolor{hellgrau}.55 (-.18) & \cellcolor{hellgrau}.81 (+/-0) & \cellcolor{hellgrau}.66 (-.11) & 
 \multirow{3}{*}{.71 (-.04)} \\ 
 & \textit{Non-Dup.}& .88 (-.10) &  .66 (-.02) & .75 (-.03)& \\
\cline{2-5}
 & \cellcolor{hellgrau}Macro avg & \cellcolor{hellgrau}.71 (-.05) & \cellcolor{hellgrau}.74 (-.01) & \cellcolor{hellgrau}\textbf{.73 (-.02)} & \\
 \hline
 \parbox[t]{2mm}{\multirow{3}{*}{\rotatebox[origin=c]{90}{\footnotesize DCCNN}}} & 
 \Dup{} & .54 (-.16) &  .70 (+/-0) & .61 (-.08) & 
 \multirow{3}{*}{.70 (+/-0)}  \\ 
 & \cellcolor{hellgrau}\textit{Non-Dup.}  & \cellcolor{hellgrau}.82 (+.12) & \cellcolor{hellgrau}.70 (+/-0) & \cellcolor{hellgrau}.75 (+.06) & \\
\cline{2-5}
 & Macro avg & .68 (-.02) & .70 (+/-) & \textbf{.69 (-.01)}  & \\
 \hline
\end{tabularx}
\end{table}

\begin{table}[h!]
\small
\setlength\tabcolsep{2.5pt}  
\centering
\captionsetup{skip=6pt}
\caption{Single Channel CNN and Dual Channel CNN model results averaged over projects for the case DOLsNL.}\label{tab:DOLvsNLResults}
\begin{tabularx}{\columnwidth}{l l L L L L} 
\hline
& \textbf{Class} & \textbf{Pre.} & \textbf{Rec.} & \textbf{F1} & \textbf{Acc.}\\
  \hline
  \multicolumn{6}{c}{On traditional \Dup{} and \NL{} test set}\\
  \hline
 \parbox[t]{2mm}{\multirow{3}{*}{\rotatebox[origin=c]{90}{\footnotesize SCCNN}}} &
 \cellcolor{hellgrau}\Dup{} & \cellcolor{hellgrau}.79 & \cellcolor{hellgrau}.72 & \cellcolor{hellgrau}.74 & 
 \multirow{3}{*}{.76}\\ 
 & \NL{} & .75 &  .81 & .77 & \\
\cline{2-5}
 & \cellcolor{hellgrau}Macro avg & \cellcolor{hellgrau}.77 & \cellcolor{hellgrau}.76 & \cellcolor{hellgrau}\textbf{.77} & \\
 \hline
 \parbox[t]{2mm}{\multirow{3}{*}{\rotatebox[origin=c]{90}{\footnotesize DCCNN}}} & 
 \Dup{} & .71 &  .66 & .68 & 
 \multirow{3}{*}{.70}  \\ 
 & \cellcolor{hellgrau}\NL{} & \cellcolor{hellgrau}.69 & \cellcolor{hellgrau}.73 & \cellcolor{hellgrau}.71 &  \\
\cline{2-5}
 & Macro avg & .70 & .70 & \textbf{.70}  & \\
  \hline
  
 \multicolumn{6}{c}{On test set with \OL{}}\\
 \hline
 \parbox[t]{2mm}{\multirow{3}{*}{\rotatebox[origin=c]{90}{\footnotesize SCCNN}}} &
 \cellcolor{hellgrau}\textit{Linked} & \cellcolor{hellgrau}.88 (+.09) & \cellcolor{hellgrau}.70 (-.02) & \cellcolor{hellgrau}0.77 (+.03) & 
 \multirow{3}{*}{.74 (-.02)} \\ 
 & \NL{}& .75 (-.15) &  .81 (+/-0) & .68 (-.09)& \\
\cline{2-5}
 & \cellcolor{hellgrau}Macro avg & \cellcolor{hellgrau}.76 (-.01) & \cellcolor{hellgrau}.74 (-.02) & \cellcolor{hellgrau}\textbf{.75 (-.02)} & \\
 \hline
 \parbox[t]{2mm}{\multirow{3}{*}{\rotatebox[origin=c]{90}{\footnotesize DCCNN}}} & 
 \textit{Linked} & .83 (+.12) &  .65 (-.01) & .73 (+.05) & 
 \multirow{3}{*}{.68 (-.02)}  \\ 
 & \cellcolor{hellgrau}\NL{}  & \cellcolor{hellgrau}.52 (-.17) & \cellcolor{hellgrau}.73 (+/-) & \cellcolor{hellgrau}.61 (-.10) & \\
\cline{2-5}
 & Macro avg & .68 (-.01) & .69 (-.01) & \textbf{.69 (-.01)}  & \\
 \hline
\end{tabularx}
\end{table}

For the second option (trained on DOLvsNL shown on Table \ref{tab:DOLvsNLResults}) the best macro F1-score, on average, is 0.75 for the SCCNN model. 
Both settings can perform rather well, as we observed differences in classification for the traditional setting between the repositories, we compared the F1-scores of the second and third setting in Table~\ref{tab:LinkedVsDupPerformance} for each repository.
We observe that the SCCNN model trained for the ``linked'' settings often performs best.
Although there are cases, namely Hyperledger, IntelDAOS, JFrog, and Spring where peformance for \Dup{} detection is better or equal.
\begin{table}[h!]
\small
\setlength\tabcolsep{2.5pt}  
\centering
\captionsetup{skip=6pt}
\caption{Comparison of performance between DvsNLOL and DOLvsNL across the  studied JIRA repositories. }\label{tab:LinkedVsDupPerformance}
\begin{tabular}{l r r r r } 
\hline
\multirow{2}{*}{\textbf{Repository}} & \multicolumn{2}{c}{\textbf{SCCNN}} &  \multicolumn{2}{c}{\textbf{DCCNN}}\\
& DvsNLOL & DOLvsNL & DvsNLOL & DOLvsNL\\
  \hline
 \cellcolor{hellgrau}Apache         & \cellcolor{hellgrau}.77 & \cellcolor{hellgrau}\textit{.85} & \cellcolor{hellgrau}.69 & \cellcolor{hellgrau}\textit{.73}\\
 Hyperledger    & \textit{.71} & .67 & \textit{.71} & .67\\
 \cellcolor{hellgrau}IntelDAOS      & \cellcolor{hellgrau}\textit{.65} & \cellcolor{hellgrau}.61 & \cellcolor{hellgrau}/ & \cellcolor{hellgrau}/ \\
 JFrog          & \textit{.66} & .64 & / & / \\
 \cellcolor{hellgrau}Jira           & \cellcolor{hellgrau}.81 & \cellcolor{hellgrau}\textit{.92} & \cellcolor{hellgrau}.76 & \cellcolor{hellgrau}\textit{.84}\\
 JiraEcosystem  & .64 & \textit{.66} & \textit{.69} & .62 \\
 \cellcolor{hellgrau}MariaDB        & \cellcolor{hellgrau}.69 & \cellcolor{hellgrau}\textit{.70} & \cellcolor{hellgrau}.65 & \cellcolor{hellgrau}.65\\
 MongoDB        & .70 & \textit{.80} & .63 & \textit{.65}\\
 \cellcolor{hellgrau}Qt             & \cellcolor{hellgrau}.77 & \cellcolor{hellgrau}\textit{.81} & \cellcolor{hellgrau}.69 & \cellcolor{hellgrau}.69\\
 RedHat         & .81 & \textit{.86} & \textit{.80} & .73\\
 \cellcolor{hellgrau}Sakai          & \cellcolor{hellgrau}.71 & \cellcolor{hellgrau}\textit{.74} & \cellcolor{hellgrau}\textit{.66} & \cellcolor{hellgrau}.65\\
 Sonatype       & .75 & \textit{.85} & / & \\
 \cellcolor{hellgrau}Spring         & \cellcolor{hellgrau}\textit{.65} & \cellcolor{hellgrau}.62 & \cellcolor{hellgrau}.62 & \cellcolor{hellgrau}.62\\
 \hline
\end{tabular}
\end{table}

Overall, we see that the DOLvsNL setting, which predicts if a mere link exists between two issues, performs the best for the SCCNN model, and overall achieves amost of the best performances across the repositories.
It seems that either the SCCNN models for duplicate detection are rather link detection models or the stakeholders' usage of links is not clear and there exist a substantial amount of mislabeling and undiscovered links.
On average, we also observe that the DCCNN model works similar for the DvsOLNL and DOLvsNL setting.
It is particularly robust, but the performance is lower in comparison to the SCCNN setting DOLvsNL.
We suspect that a Dual-Channel architecture might be better at distinguishing link types from each other.
Additionally, the performance varies strongly with the repository, which suggests that we might need to adapt detection approaches to the characteristics of the repository.

Overall, none of the settings on the JIRA datasets achieved a similarly high accuracy in comparison to the Bugzilla datasets~\citep{He:ICPC:2020}, which was above 0.94 for Open Office, Eclipse, and NetBeans.

\section{Discussion}\label{sec:discussion}
We summarize the main findings and potential implications for research and practice. Then we briefly highlight potential limitations and threats to validity.

\subsection{Summary of Findings}
Bugzilla has three default link types and users can also link issues via comments.
In GitHub referencing another issues or pull requests in the comments is the main way to establish an issue link.
In contrast, JIRA dedicates specific sections for creating default and custom link types.
The flexibility of issue link creation in JIRA seems to allow stakeholders create a plethora of link types and customize the ITS to the specific needs of their projects.

Our manual analysis of the different link types has resulted in a link categorisation including \Rel{}, \Dup{}, \Com{}, \Tem{}, and \Wor{} -- a first step towards a universal taxonomy of linking in ITS.  
\Rel{} links are used to point to useful information included in other issues. 
This link category could be used as a reminder to specify another concrete link type, which could be investigated by looking at the history of issues.
An example is issue ZOOKEEPER-3920 in Apache. A \Rel{} link connects this issue to ZOOKEEPER-3466 and ZOOKEEPER-3828, while the comments discuss if this should be a \Dup{} link.
\Com{} links are used to manage issues and the corresponding workload as they are used to create a work breakdown.
\Dup{} links are used to indicate duplicate issues, which are important to identify to prevent duplication of effort.
Finding duplicate issues can provide additional context information which is useful to resolve the issue~\citep{Zimmermann:TSE:2010}.
\Tem{} links are useful for planning releases as they indicate the dependencies between issues.
\Wor{} links are perhaps useful for quality assurance, validation, and testing.
These categories were regularly used in all analyzed JIRA repositories except for 4 cases out of 75 usages  (5 categories x 15 repositories).

Looking at the structural differences and similarities of the graphs from the different link categories, we observed that the category \Rel{} makes up the most transitive graphs among all link type categories, while \Dup{} and \Tem{} should be transitive by definition. One possible reason could be that Stakeholders favor more efficient representations of the issue graph. 
Inferrable transitive links of a link type category tend not to be included explicitly.
For example, if issue A is a component of issue B, which is again a component of issue C, then the \Com{} link between issues A and C tends not to be explicitly created  by stakeholders.
One possible explanation is that stakeholders aim to avoid an information overload on the single issue pages.
In general, we think that the analysis of structural differences and graph metrics can be leveraged to find inconsistencies in the issue graph or could be refined to track the health of an ITS.
A high number of isolated issues could, for instance, indicate a lot of unknown or undocumented links.

Lastly, the results of RQ2 show that current duplicate detection techniques are, not only significantly less accurate when applied on JIRA, but are also unable to correctly distinguish between duplicate and other link types. This limits their applicability in practice.
The DCCNN approach by He et al.~\citep{He:ICPC:2020} was more robust than the SCCNN approach.
We also found that the SCCNN approach performs the best when used as a general link detection with a few exceptions. This should be further investigated. 
Generally, there is a significant room for improving link prediction models applied on a realistic dataset with heterogeneous link types: both for predicting duplicates and other links. 


\subsection{Implications for Practitioners}
Mining the issue links in an issue tracker and evaluating fitting metrics on the underlying issue graph could help overviewing the issue repository and potentially \textbf{monitoring ''ITS health''} for supporting project management and release planning activities. 
Stakeholders could decide about the characteristics of ''healthy issue graph''  and their priorities.
A healthy issue graph could, e.g., contain only minimal needed edges with no transitivity to avoid information overload, correct direction of issue links, correct link types, and no sets of issues where the links are conflicting to each other (as issues which create a directed cycle with \dep{} edges)~\citep{tiihonen:2019:coping}.
Another unwanted structure in an issue graph could be a set of issues with many (unnecessary) links and a single link leading to another set of issues with many (unnecessary) links.
The single link that connects these issue clusters can be missed due to information overload.

We showcased a few \textbf{possible metrics} and their interpretations.
For instance, \Com{} showed a hierarchical, tree-like structure, but not in all cases. These exceptions could be analyzed to find common errors or interesting insights for the corresponding repository.
\Dup{} components that are not star-shaped might not be optimal, as one original bug should exist. 
Duplicates of duplicates might contain important information, which is not visible to the developer when they view the original issue.
By looking in which circumstances a transitive link was deemed important enough to be made explicit, these could point to issues that have higher importance.
A dashboard calculating such metrics per individual link type could make it easier to identify outliers issue clusters which then can be investigated and if necessary, cleaned.

As link types and definitions of custom types can change over time, it might also be useful to monitor the usage frequency over time. \textbf{Unused link types} can be rechecked and eventually removed in order to avoid confusing users with too many possible link types.
Information overload can also be reduced by analyzing and potentially consolidating certain link types of the same category. Our categorization scheme and the overall prevalence and complementary usage across the repositories can provide a concrete guidance for practitioners. 
Issues with many links or multiple link types might also indicate that an issue is too big and need to be broken down into multiple parts~\citep{Nicholson:AIRE_W:2020}.

\subsection{Implications for Researchers}
It is important for software engineering research to be applicable in real-world scenarios as one of its main aims is to support the work of software practitioners. 
Better \textbf{understanding how stakeholders use links} to organize project issues should thus improve the applicability of research on issue tracking, particularly machine learning models to predict related -- e.g. duplicated -- issues. 
Our experimental results show that state-of-the-art deep learning approaches for duplicate prediction are unable to precisely classify \OL{} as Non-Duplicates in most JIRA repositories.
Future research should further evaluate this across other datasets, as Bugzilla also includes other link types besides \dup{}.

Indeed, we think that our \textbf{model evaluation strategy} could also inspire researchers to test other machine learning models (particularly binary classifiers) on  more realistic settings, by trying to confuse them. The idea is simple: negative classes should explicitly include similar items to the positive class, which are still negative though. This enables checking to what extent a model gets confused even when the "new" negative class is represented in the training data. In our research we included instances of \OL{} in the negative class \NL.

Our results suggest that Dual-Channel approaches (which embed both linked issues and thus the links itself) seems more robust when presented with \OL{}. 
Single-Channel approaches seem able to detect links instead of duplicates as this model performed the best for this task on our dataset.
A two-step approach, first detecting the existence of a link and then the specific link type, might be a reasonable and effective trade-off for detecting other link types besides duplicates. 


We also observed significant differences in the prediction performance between the various JIRA repositories.
This suggests that the characteristics of the underlying repository should be taken into account: including for instance the exact link semantic shared across a project members as well as the quality of the linking (i.e. probability of human mistakes when setting links and link types). 
Another interesting finding is that link transitivity is barely explicitly documented in JIRA. 
Stakeholders seem to rely on implicit transitive links -- a potentially useful information to down-prioritizing implicit links in link recommendations. 
Cases where stakeholders make implicit transitivity explicit are worth researching  to understand specific issue context or particular preferences of the stakeholders.

Overall, our results call for more intensive replication studies with known duplicate detection models on different ITS and repositories. 
Such studies should inform what type of link detection and what model architecture work robustly on what ITS and what links.
Furthermore, redoing the analysis on link types instead of the categories can reveal more fine-grained differences.
More complex graph metrics can be designed or ''borrowed'' from social network research to find more insights into link usage in ITS~\citep{Hansen2020}.

\subsection{Threats to Validity}\label{sec:threats}
During link type categorization, we observed that not all link types are unequivocally categorizable and that interviews with key stakeholders would be needed to exactly understand the link usage within the repository. 
We tried to reduce this threat by looking at and discussing multiple examples for each link type and repository.

The data is manually created and labeled by stakeholders.
Therefore we cannot rule out the possibility of undiscovered and missing links as well as stakeholders not always agreeing on the way to use the link types.
For instance, issue CONJ-740 in MariaDB is documented as the cause for issue CONJ-664, but a  comment on this issue states that it is a duplicate and the issue is closed as such.

Furthermore, we were not able to access private issues from the investigated repositories.
As a result, our dataset might not be representative of the complete population of issues and their links within each repository.
Additionally, some repositories (e.g.: RedHat) tend to create custom fields for links instead of a custom link type, which are usually in the format ``customfield\_number''.
We looked through many issues from different repositories to find and identify all such fields that link different issues together. 

We used a single-channel approach to represent the other SotA models and replicated the exact dual-channel approach as it was reported to work best for duplicate detection.
All duplicate detection models leverage that duplicate issues should be very similar and non-linked issues very dissimilar and only differ in their layers to detect similarity.
But, in practice, issues can have other types of links and have varying degrees of similarity.
All models that only leverage similarity will then struggle to correctly asses \OL{} as Non-Duplicates.
Some approaches, as the dual-channel approach, can be more robust to this problem.
Furthermore, we did not test the models on Bugzilla and GitHub issues, but we report results on multiple JIRA.
Thus, generalizability to other ITS might be limited.

\section{Related Work}

\label{sec:rel-work}
Issue tracking systems have originally been geared toward collaborative bug fixing.
Early research such as the work by Zimmermann et al.~\citep{Zimmermann:TSE:2010} closely analyzed ITS from the perspective of bug report quality.
They showed that duplicate bug reports are one of the most common problems in ITS, after bug reports containing missing or erroneous information.
With the expansion of ITS to include issue types other than bug reports, further works emerged focusing on issue type classification.
Herzig et al.~\citep{Herzig:ICSE:2013} for instance found that a substantial percentage of bug reports (33.8\%) from a sample of 7000 were misclassified as bugs instead of feature requests.
The authors argue that such misclassifications introduce bias to bug prediction models.
Subsequent works have since used information retrieval and machine learning techniques leveraging issue description text and metadata to solve the bug report classification problem~\citep{Terdchanakul:ICSME:2017}~\citep{zhou:JSE:2016}.
In requirements engineering research, on the other hand, focus has been more directed towards the detection of feature requests  ~\citep{Merten:RE:2016}\citep{Qiang:ESEM:2017}.

As ITS platforms mature,
they face challenges related to repository growth and evolution.
A case study conducted by Fucci et al.~\citep{Fucci:ESEM:2018} among stakeholders of a software company using JIRA to document requirements found that information overload is one of the biggest challenges faced by stakeholders.
Interviewees of the study expressed the need for a requirements dependency identification functionality to reduce the overhead of discovering and documenting dependencies manually.
Other challenges concern ITS platforms being historically driven by bug reports and focusing less on features and requirements.

Furthermore, the usage of ITS requires continuous maintenance and cleanup of the issues.
Projects that use ITS often undertake triaging~\citep{Anvik:ICSE:2006}, whereby issues are revised, prioritized, and duplicate issues are identified.
Several works have proposed approaches to facilitate triaging.
Lamkanfi et al. focused on mining and predicting the severity of Bugzilla issues~\citep{Lamkanfi:MSR:2010}\citep{Lamkanfi:ICMSE:2011}, which directly influences their prioritization.
Xuan et al. \citep{Xuan:ICSE:2012} leveraged developer social networks to understand developer prioritization and used the results for triaging, severity prediction and reopening detection, which greatly facilitates processes such as release planning.
Jeong et al.~\citep{Jeong:FSE:2009} used Markov chains to capture the bug tossing history in developer networks and team structures to improve triaging.

Duplicate issue detection in particular is a tedious task when manually curating a repository's ITS. 
However, the problem lends itself perfectly to automation using machine learning due to the high assumed similarity of duplicate pairs.
The overwhelming majority of scientific studies have therefore focused on duplicate link detection ~\citep{WANG:ICSE:08}\citep{Bettenburg:ICSM:2008}\citep{Lazar:MSR:2014} relegating the more general problem of link type detection away from the center of attention.
Deshmukh et al.~\citep{Deshmukh:ICSME:2017} proposed a single-channel siamese network approach with triplet loss using a combination of CNNs and LSTMs to detect duplicate bug reports. 
They report an accuracy close to 90\% and recall rate close to 80\%.
He et al.~\citep{He:ICPC:2020} proposed a dual-channel approach and achieved an accuracy of up to 97\%.
Rocha et al.~\citep{Rocha:ACCESS:2021} created a model using all ``Duplicate'' issues as different descriptions of the same issue and split the training and test along cluster boundaries. 
All three works~\citep{Deshmukh:ICSME:2017, He:ICPC:2020, Rocha:ACCESS:2021} use the dataset provided by Lazar et al.~\citep{Lazar:MSR:2014}, containing data mined from the four open source Bugzilla systems: Eclipse, Mozilla, NetBeans, and OpenOffice.


Despite \Dup{} being the most widely researched link type, other link types have also been considered for analysis although much less extensively.
For instance, Thompson et al.~\citep{Thompson:MSR:2016} studied three open source systems and analyzed how software developers use work breakdown relationships between issues in JIRA. 
They observed little consistency in the types and naming of the supported relationships. 
Merten et al.~\citep{Merten:REFSQ:2016} studied links from the traceability perspective and reported a poor applicability of traceability approaches in ITS due to poor quality.
This represents an important motivation for the use of metrics in our work to help identify link inconsistencies for better predictions.
Tomova et al.~\citep{Thompson:MSR:2016} studied seven open source systems and reported that the rationale behind the choice of a specific link type is not always obvious.
The authors found that while the \clo{} link is indicative of textual similarity, issues linked through a \rel~link presented varying degrees of textual similarity and thus require further contextual information to be accurately identified.
Li et al.~\citep{Li:APSEC:2018} examined the issue linking practices in GitHub and extracted emerging linking patterns. 
In their work, the authors categorized the link types into 6 link type categories, namely: ``Dependent'', ``Duplicate'', ``Relevant'', ``Referenced'', ``Fixed'', ``Enhanced'', all the rarer link types were assigned the category ``Other''. 
They furthermore found patterns facilitating the automatic classification of the link types.
For instance, they discovered that ``Referenced'' links  usually refer to historic comments with important knowledge and that ``Duplicate'' links are usually marked within the day.

Deshpande et al.~\citep{Deshpande:RE:2020} examined ``Requires'' and ``Refines'' links.  The authors extracted dependencies by integrating an  active learning approach with ontology-based retrieval on two industrial datasets achieving an F1-score of at least 75\% in both training sets.
 Cheng et al.~\citep{Cheng:COMPSAC:2020} examined \blo{} links.
They used the repositories mined by Lazar et al.~\citep{Lazar:MSR:2014} and predicted the \blo{} link type with an F1-Score of 81\% and AUC of 97.5\%.

As observed by Bertram et al.\citep{Bertram:CSCW:2010}, ITS serve as a focal point for communication, especially through the use of issue comments.
Linking issues to each other is also a form of communication, as it serves to link two discussion threads.
Additionally, stakeholders often refer to other issues in issue discussions, as reported by Arya et al.~\citep{Arya:ICSE:2019}. 
The authors categorized and predicted the issue comments to improve communication and collaboration processes in GitHub Issues.
Communication gaps in large scale software development lead to failure to meet customers' expectations, quality issues, and wasted effort~\citep{Bjarnason:RE:2011}.
Issue link research can improve other areas such as Requirements Engineering through early failure prediction of feature requests~\citep{Fitzgerald:RE:2011} as linked issues provide more context to be leveraged by a classifier.
Seiler et al.~\citep{Seiler:REFSQ:2017} found that 
insufficient traceability and fragmentation of feature knowledge are major practical  problems. 
Link detection can help address these problems.
Qiang et al.~\citep{Qiang:ESEM:2017} found that issue report classification based on text mining is difficult to adopt as contributors only write short texts. 
Other information such as structural data as well as context provided by links can help addressing this challenge.

\section{Conclusion}
\label{sec:conclusion}
We studied 30 distinct link types in 15 large open source organizations and identified an overarching categorization to help the semantic clutter of issue linking. We reported on several expected and unexpected trends concerning how the link types and the categories are used in practice: particularly in term of prevalence and the structural properties of the underlying issue graphs. 
Furthermore, we examined how Single-Channel and Dual-Channel deep-learning approaches to detect the specific duplication links perform with different training data. Both approaches seem unable to correctly recognize \OL{} as non-duplicates.
Future duplicate detection models needs to consider other link types for a higher prediction reliability and thus better applicability in practice.

On one hand, link detection approaches can create two-step models that are able to distinguish the different link types. 
First, the existence of a link is predicted and then, with a second model, its type.
Conducting a more detailed error analysis of existing duplicate detection models on various datasets from different ITS will lead to a better understanding of how to reduce the models uncertainty.
On the other hand, our observations of the usage of link types can inform how to further enhance the outputs. 
For instance, by analyzing the transitivity of a link type in a specific repository, we can remove transitive links which are already implicitly contained in the system.
Our link analysis metrics can be used to monitor ITS usage  and explain the performance of a specific model architecture.
Overall, our results open up a new avenue for link type analysis and detection beyond duplicates. 

\section*{Acknowledgment}
We thank Lloyd Montgomery for collecting the dataset.
This work has been partly conducted within the Horizon 2020 project OpenReq, which is supported by the European Union under the Grant Nr. 732463. The second author is funded by the German Science Foundation DFG (MA 6149). 

\bibliographystyle{ACM-Reference-Format}
\bibliography{references}

\end{document}